\documentclass[journal]{IEEEtran}
\usepackage{cite}
\usepackage{amsmath,amssymb,amsfonts}

\usepackage{algorithm} 
\usepackage{color,xcolor}
\usepackage{subfigure}
\usepackage{textcomp}
\usepackage{bm}
\usepackage{amsmath,amsfonts}
\usepackage{algorithmic}
\usepackage{array}
\usepackage[caption=false,font=normalsize,labelfont=sf,textfont=sf]{subfig}
\usepackage{textcomp}
\usepackage{stfloats}
\usepackage{url}
\usepackage{verbatim}
\usepackage{graphicx}
\usepackage{tabularx}
\usepackage{caption}
\usepackage{array}
\def\BibTeX{{\rm B\kern-.05em{\sc i\kern-.025em b}\kern-.08em
    T\kern-.1667em\lower.7ex\hbox{E}\kern-.125emX}}
\usepackage{balance}
\begin{document}
\title{Robust Computation Offloading and Trajectory Optimization for Multi-UAV-Assisted MEC: A Multi-Agent DRL Approach}

\author{Bin Li, Rongrong Yang, Lei Liu~\IEEEmembership{Member,~IEEE}, Junyi Wang,~\IEEEmembership{Member,~IEEE}, Ning Zhang,~\IEEEmembership{Senior Member,~IEEE}, and Mianxiong Dong,~\IEEEmembership{Senior Member,~IEEE}

\thanks{B. Li and R. Yang are with the School of Computer and Software, Jiangsu Collaborative Innovation Center of Atmospheric Environment and Equipment Technology (CICAEET), Nanjing University of Information Science and Technology, Nanjing 210044, China (bin.li@nuist.edu.cn; 202212210020@nuist.edu.cn).}

\thanks{L. Liu is with the Guangzhou Institute of Technology, Xidian University, Guangzhou 510555, China (e-mail: tianjiaoliulei@163.com).}

\thanks{J. Wang is with the School of Information and Communication, Guilin University of Electronic Technology, Guilin 541004, China (e-mail: wangjy@guet.edu.cn).}

\thanks{N. Zhang is with the Department of Electrical and Computer Engineering, University of Windsor, Windsor, ON N9B 3P4, Canada (e-mail: ning.zhang@uwindsor.ca).}

\thanks{M. Dong is with the Department of Sciences and Informatics, Muroran Institute of Technology, Muroran, Japan (e-mail: mx.dong@csse.muroran-it.ac.jp).}
}

\maketitle

\begin{abstract}
    For multiple Unmanned-Aerial-Vehicles (UAVs) assisted Mobile Edge Computing (MEC) networks, we study the problem of combined computation and communication for user equipments deployed with multi-type tasks.
    Specifically, we consider that the MEC network encompasses both communication and computation uncertainties, where the partial channel state information and the inaccurate estimation of task complexity are only available.
    We introduce a robust design accounting for these uncertainties and minimize the total weighted energy consumption by jointly optimizing UAV trajectory, task partition, as well as the computation and communication resource allocation in the multi-UAV scenario.
    The formulated problem is challenging to solve with the coupled optimization variables and the high uncertainties.
    To overcome this issue, we reformulate a multi-agent Markov decision process and propose a multi-agent proximal policy optimization with Beta distribution framework to achieve a flexible learning policy.
    Numerical results demonstrate the effectiveness and robustness of the proposed algorithm for the multi-UAV-assisted MEC network, which outperforms the representative benchmarks of the deep reinforcement learning and heuristic algorithms.
\end{abstract}

\section{Introduction}
As the Internet of Things (IoT) era continues to advance, modern society is becoming increasingly reliant on IoT technology \cite{EItCo}. 
This has led to the creation of the massive data at the edge nodes of the networks.
How to deal with these data quickly and effectively has become a significant problem, which is worthy of consideration.
Mobile Edge Computing (MEC) as a new computing paradigm has been introduced, where nearby servers are utilized as edge clouds to provide User Equipments (UEs) with powerful cloud computing capabilities while significantly reducing the time delay of computation offloading \cite{tEIVi}.

Nevertheless, serving intensive tasks in remote areas is very challenging due to poor communication conditions and unstable MEC environments \cite{ECAAF}.
Meanwhile, in some hotspot areas, when a large number of UEs require computation-intensive services simultaneously, the limited computation and storage resources pose a formidable challenge for MEC servers in guaranteeing the satisfactory user experience.
To tackle these issues, flexible location deployment of edge servers is essential.
Hence, Unmanned Aerial Vehicle (UAV) has been used as a popular platform for the MEC network owing to its superior ability of high mobility and coverage enhancement, where UAV edge server can assist in the remote areas and alleviate congestion in hotspot areas to ensure the high-quality computing services.

Although UAV-assisted MEC network has attracted enormous research interests, it still faces many uncertainties in practice.
Firstly, computation offloading is subject to unpredictable delivery time and packet loss rate owing to the heterogeneous MEC networks, which in turn leads to the unreliable edge computing nodes\cite{RSPfE}.
In addition, offloading decision is usually dependent on the accurate Channel State Information (CSI), it is quite difficult to obtain \cite{aSoME}.
The time-varying channels based on precise CSI bring uncertainties with respect to the computation offloading rate, thus increasing the offloading delay.
Moreover, in practical applications, the task complexity of the computing tasks can only be obtained exactly after the task is completed.
As a result, there may be unexpected delays in the calculation time, even the system fails to return the results to mobile devices in a timely manner.
Under such conditions, robust design plays a crucial role in providing worst-case performance guarantees against possible failures.

A single UAV cannot efficiently serve a large number of UEs owing to the restricted coverage and computing capability, which in nature spurs our exploration of multi-UAV cooperation. Also, edge networks may experience uncertainties in both communication and computation, but previous studies mainly focused on individual robustness \cite{Robust0}. 
To address the above problem, we propose a robust offloading scheme in the MEC network where multiple UAVs collaborate to serve numerous UEs. 
We jointly consider the imperfect CSI between UAVs and UEs, as well as the uncertainties related to the task complexity. Our scheme aims to enhance the robustness of the system while minimizing the weighted energy consumption.
The main technical contributions from this paper are summarized as follows:
\begin{enumerate}
    \item We investigate the computation uncertainties and communication uncertainties in a multi-UAV-assisted MEC network. To ensure the robustness of the computation offloading process, we formulate a problem for minimizing the total energy consumption of the system through the joint optimization of UAV trajectories, selection factors between UAVs and UEs, task partition, as well as the communication and computation resources between UAVs and UEs.
    \item The formulated problem involves tightly coupled optimization variables and the uncertainty constraints, posing a challenge to find the global optimal solution. To this end, we resort to the deep reinforcement learning and propose a Multi-Agent Proximal Policy Optimization (MAPPO) algorithm. Additionally, to eliminate the boundary effects caused by Gaussian distribution in the original MAPPO algorithm, we utilize the Beta distribution in the output of the actor network.
    \item We evaluate the complexity of the MAPPO with Beta distribution (b-MAPPO) algorithm, and demonstrate its convergence and robustness in guaranteeing the energy consumption minimization under the bounded estimation errors through the numerical results.
\end{enumerate}

The rest of this paper is organized as follows. Related works are reviewed in Section II.  In Section III, we introduce the system model. Then, we propose the b-MAPPO framework and analyze its complexity in Section IV.  Section V provides extensive simulations to verify the robustness and effectiveness of the proposed algorithm.  Finally, Section VI makes a conclusion.

\section{Related Work}
Several studies try to tackle the related issues for computation offloading in MEC networks, such as service delay \cite{MEC2,MEC7}, bandwidth \cite{MEC1}, power consumption \cite{MEC3} or balance time and energy consumption \cite{MEC4}.
The conventional studies on MEC networks mainly focuse on fixed base stations deployed on the ground, but lack service flexibility. 
To address this limitation, UAV is introduced into the MEC networks \cite{UAV1,UAV2,UAV7,UAV6} to enhance the user experience in remote areas or hotspot areas.
By taking into account the dependencies between various tasks, the authors of \cite{UAV1} investigated the energy consumption minimization problem by jointly optimizing the resource allocation, UAV trajectory, and offloading decision.
In \cite{UAV2}, the authors conducted the research on minimizing the average energy consumption for multi-UAV cellular-connected MEC networks.
In \cite{UAV7}, the authors designed a two-layer optimization approach which jointly optimizes bit allocation, UAV trajectory, and UAV task scheduling with the objective of minimizing the energy consumption for UEs.
In \cite{UAV6}, a multi-UAV-assisted MEC framework was used, where a UAV is controlled by a dedicated agent to jointly optimize the trajectory and offloading decisions of the UAV.
In \cite{UAV4}, the authors jointly optimized terminal device scheduling, time slot size, and UAV trajectories to minimize the completion time of the tasks under the considering of both partial offloading and binary offloading modes.
The authors of \cite{UAV5} considered a scenario with multi-edge-cloud and multi-UAV and employed Multi-Agent Deep Reinforcement Learning (MADRL) to solve the computation offloading problem with the aim of minimizing the sum cost.
The work in \cite{UAV3} took into account the coordination advantage of multiple UAVs in a fleeted way and maximized the system energy efficiency via alternating direction method of multipliers algorithm and Lyapunov optimization. 
Although the above researches have well applied UAVs into MEC to enhance the network flexibility, they did not consider the robustness problem.

For practical MEC networks, the availability of CSI and the task complexity are one of the utmost significant problems in implementing the computation offloading.  
As a result, robust design is critical to offer performance guarantees for optimization problems with the uncertainties.
Generally, the robust design can be classified into three types: \textit{scheduling robustness design}, \textit{channel robustness design}, and \textit{computation robustness design}. 
For \textit{scheduling robustness design},
the authors of \cite{Robust1} formulated a robust task scheduling problem in the case of uncertain offloading failure with the purpose of minimizing the latency.
The authors of \cite{Robust2} proposed a robust anti-edge server fault task offloading scheme to overcome the dynamics of edge servers.
For \textit{channel robustness design},
the authors of \cite{Robust3} presented a hybrid offloading scheme with backscatter communication under imperfect CSI with the aim of minimizing the end-to-end system latency.
In \cite{Robust4}, the authors considered a robust offloading strategy against realistic channel estimation errors in fog-IoT systems and minimized the power consumption of UEs with the latency requirements.
In \textit{computation robustness design},
the authors in \cite{Robust5} investigated the fog radio access network in which  the knowledge of computation provision with bounded perturbations is inaccurate and developed a computation offloading mechanism with the goal of minimizing the UEs' energy  consumption.
In \cite{Robust6}, the authors focused on the demand uncertainty with a single cache-enabled UAV and minimized the delay brought by the UAV-assisted caching by jointly optimizing the trajectory and caching of the UAV.
In \cite{Robust7}, the authors minimized the maximum system delay in a multi-task MEC network with a base station by taking into account the communication and computation uncertainties.
In view of prior work, there is little research focusing on the robust computation offloading in UAV-assisted MEC networks.
Against this background, we investigate the communication and computation uncertainties in a multi-UAV-assisted MEC network.
Unlike the existing work \cite{Robust7}, we take the collaboration between UAVs into consideration to provide services for UEs more flexibly.

\begin{figure}[t]
    \centering
    \includegraphics[width=\columnwidth]{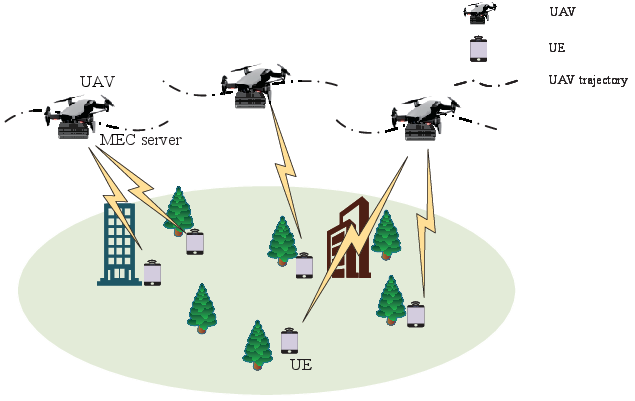}
    \caption{System model of the proposed multi-UAV-assisted MEC network.}
    \label{fig:sys-model}
\end{figure}
\section{System Model and Problem Formulation}\label{s:sys}
We investigate a multi-UAV-assisted MEC network as shown in Fig. \ref{fig:sys-model}, which is composed of $M$ UAVs and $K$ UEs.
Note that UAVs consist of a Uniform Planar Array (UPA) with $At=A_x \times A_y$ antennas and UEs are equipped with
one single antenna each. To facilitate expression and analysis, we define the collection of indexes for UAVs as $\forall m \in \mathcal{M}\triangleq\{1,2,\cdots,M\}$,
the collection of indexes for UEs as $\forall k \in \mathcal{K}\triangleq\{1,2,\cdots,K\}$, 
and the collection of indexes for time slots as $\forall n \in \mathcal{N}\triangleq\{1,2,\cdots,N\}$.
And we define UAVs' flight period as $T=N\delta$, in which $\delta$ is the time duration of the time slot.
Assume that the resource-intensive computation tasks are generated in each time slot for each UE. These tasks need to be completed during a given time deadline.
We define the task of UE $k$ during the $n$-th time slot as $D_k\left[ n \right]$. 
Considering the limited resources of UEs and based on the position information of UAVs, each UE can select a UAV for computation offloading. The matching factor between UAVs and UEs can be represented as
\begin{align}
    &\sum\limits_{m=1}^{M}{\alpha_{k,m} \le 1},\forall k \in \mathcal{K},\\
    &\alpha_{k,m} \in \{0,1\}, \forall k \in \mathcal{K}, m \in \mathcal{M},
\end{align}
where $\alpha_{k,m}=1$ if UAV $m$ is chosen to offload the tasks by UE $k$, and $\alpha_{k,m}=0$ otherwise.

\subsection{UAV Movement Model}
Without loss of generality, we will use the Cartesian coordinate system.
The fixed position of UE $k$ can be represented as ${\bf u}_k = \left(x_k, y_k \right)^{\rm T}$, while ${\bf q}_m\left[ n \right] = \left(x_m\left[ n \right], y_m\left[ n \right] \right)^{\rm T}$ represents the horizontal coordinate of UAV $m$ during the $n$-th time slot.
Assume that UAVs  maintain a constant altitude $H$ above the ground such that they can avoid frequent ascent and descent to save energy. 

To avoid collisions and conflicts, the UAVs need to consider the positions and movements of other UAVs while planning their own paths, thus ensuring effective task execution. Therefore, the transformations of UAV positions between different time slots, which are related to flight speed ${\bf v}_m\left[ n \right]$ and acceleration ${\bf a}_m \left[ n \right]$, should satisfy the following constraints
\begin{align}
    &{\bf q}_m \left[ n+1 \right] = {\bf q}_m \left[ n \right] +{\bf v}_m \left[ n \right]\delta + \frac{1}{2} {\bf a}_m \left[ n \right]\delta^2,\\
    & \Vert {\bf q}_i \left[ n \right] -{\bf q}_j \left[ n \right] \Vert^2 \ge d_{\rm {dim}}^2,
\end{align}
where $d_{\rm{dim}}$ is the minimum safe distance when UAVs flying.
And $\Vert {\bf a}_m \left[ n \right] \Vert$ is given by 
\begin{equation}
    \Vert {\bf a}_m \left[ n \right] \Vert = \frac{\Vert {\bf v}_m\left[ n+1\right]\Vert - \Vert {\bf v}_m\left[ n\right]\Vert}{\delta}.
\end{equation}

When a UAV flies, its propulsion power consumption $p_m^{\rm{fly}}\left[ n \right]$ is modeled as \cite{channel2}
\begin{align}
    p_m^{\rm{fly}}\left[ n \right] =&\nonumber \frac{1}{2} d_0 \rho g A_0 \Vert {\bf v}_m\left[ n \right] \Vert^3 + P_1\left( 1+ \frac{3\Vert {\bf v}_m \left[ n \right] \Vert^2}{U_{\rm{tip}}^2}\right)\\
    &+P_2 \left( \sqrt{1+ \frac{\Vert {\bf v}_m \left[ n \right] \Vert^4}{4v_0^4}} - \frac{\Vert {\bf v}_m \left[ n \right] \Vert^2}{2v_0^2} \right)^{\frac{1}{2}},
\end{align} 
where $P_1$ is the power of UAV's blade, $P_2$ is the induced power during hovering, $v_0$ is the mean velocity of rotors and $\rho$ is  the air density.
$U_{\rm{tip}}$ is the blade's tip speed, $d_0$ denotes the fuselage drag ratio, $A_0$ represents the area of rotors and $g$ means the rotor solidity.

Consequently, the flying energy consumption of UAV $m$ during the $n$-th time slot is calculated as $E_m^{\rm {fly}} \left[ n \right] = p_m^{\rm{fly}}\left[ n\right] \delta$.

The total energy consumption of flight during the $n$-th time slot is written as 
\begin{equation}
    E_{\rm{fly}} \left[ n \right] = \sum \limits_{m=1}^M {E_m^{\rm{fly}} \left[ n \right]}.
\end{equation}

\subsection{Communication Model}
In the complex environment with obstacles like buildings and trees, the Line-of-Sight (LoS) links between UEs and UAVs are obstructed.
Consequently, the channels between UAVs and UEs exhibit Rayleigh block fading, which encompasses both Non-Line-of-Sight (NLoS) and LoS components.
The estimated CSI between UAV $m$ and UE $k$ during the $n$-th time slot is calculated as \cite{channel1}
\begin{equation}
    {\bf{\hat{h}}}_{k,m}\left[ n \right] = \sqrt{\rho d_{k,m}^{-\beta} \left[ n \right]} \left( \sqrt{\frac{\varsigma}{\varsigma+1}} {\bf{\bar h}}_{k,m}^L \left[ n \right] + \sqrt{\frac{1}{1+\varsigma}} {\bf{\tilde{h}}}_{k,m}^N \left[n \right] \right),
\end{equation}
where $\beta$ denotes the path-loss exponent, $d_{k,m}\left[ n \right]$ denotes the distance between UAV $m$ and UE $k$ during the $n$-th time slot, and $\varsigma$ denotes the Rician factor.
${\bf{\bar h}}_{k,m}^L \left[ n \right] \in \mathbb{C}^{At \times 1}$ is the LoS component from UAV $m$ to UE $k$ during the $n$-th time slot, which is denoted as
\begin{align}
    {\bf{\bar h}}_{k,m}^L \left[ n \right] = & \nonumber \left( 1,\cdots, e^{-j \frac{2 \pi b f_c}{c} \sin {\bar \omega}_{k,m}\left[ n \right] \left( a_x -1 \right) \cos\phi_{k,m}\left[ n \right]},\right.\\
     &\left. \nonumber \cdots,e^{-j \frac{2 \pi b f_c}{c} \sin {\bar \omega}_{k,m}\left[ n \right] \left( A_x -1 \right) \cos\phi_{k,m}\left[ n \right]} \right)\\
     & \nonumber \otimes \left( 1,\cdots, e^{-j \frac{2 \pi b f_c}{c} \sin {\bar \omega}_{k,m}\left[ n \right] \left( a_y -1 \right) \sin\phi_{k,m}\left[ n \right]},\right.\\
     & \left. \nonumber \cdots,e^{-j \frac{2 \pi b f_c}{c} \sin {\bar \omega}_{k,m}\left[ n \right] \left( A_y -1 \right) \sin\phi_{k,m}\left[ n \right]} \right),\\
\end{align}
where we define $b$ as the antenna inter-element spacing, and $c$ as the UAVs' speed when they fly.
The parameter $f_c$ represents the center frequency of the information carrier while $a_x$ and $a_y$ denote the row and column indices of UPA.
We define the horizontal angle of departure (AoD) and the vertical AoD from UAV $m$ to UE $k$ during the $n$-th time slot as ${\phi_{k,m}\left[ n \right]}$ and ${\bar \omega}_{k,m}\left[ n \right]$, respectively.
Particularly, the AoDs can be formulated as \cite{channel2}
\begin{align}
    & {\bar \omega}_{k,m}\left[ n \right] = \arcsin \frac{H}{\sqrt{\Vert {\bf q}_m \left[ n \right] - {\bf u}_k \Vert^2 + H^2}},\\
    & \phi_{k,m}\left[ n \right] = \arccos \frac{y_m\left[ n \right] - y_k}{\Vert {\bf q}_m \left[ n \right] - {\bf u}_k \Vert}.
\end{align}

Besides, the NLoS component $ {\bf{\tilde{h}}}_{k,m}^N \in \mathbb{C}^{At \times 1} $ is given by a complex Gaussian distributed with zero mean and unit variance, i.e., ${\bf{\tilde{h}}}_{k,m}^N \sim\mathcal{CN}({\bf 0},{\bf I})$.  

In practical MEC networks, acquiring perfect CSI is challenging due to limitations such as feedback, quantization errors, and channel estimation.
To account for these uncertainties, a commonly used approach is to employ a deterministic imperfect channel model \cite{Robust7}, which can be written as
\begin{equation}
    {\bf h}_{k,m} \left[ n \right] = {\bf {\hat{h}}}_{k,m} \left[ n \right] + \Delta {\bf h}_{k,m} \left[ n \right], \Vert \Delta {\bf h}_{k,m} \left[ n \right] \Vert \le \varepsilon_{k,m},
\end{equation}
in which ${\bf {\hat{h}}}_{k,m} \left[ n \right]$ represents the estimated CSI and $\Delta {\bf h}_{k,m} \left[ n \right]$ represents the channel error vector, subject to the constraint that the norm of $\Delta {\bf h}_{k,m} \left[ n \right]$ falls within a given radius $\varepsilon_{k,m}$.

It is desirable to utilize UAVs for edge computing by offloading tasks to them.
After the tasks are finished, the computed results are transmitted to UEs through the downlink.
To accomplish this, we begin by creating a transmit signal of the task $D_{k}\left[ n \right]$ as $x_k\left[ n \right] = \sqrt{p_k\left[ n \right]} s_k\left[ n \right]$,
in which $p_k\left[ n \right]$ is the transmission power of UE $k$ during the $n$-th time slot. $s_k \left[ n \right]$ represents the unit-norm signal for the task $D_k\left[ n \right]$, which is distributed according to the Gaussian distribution. 
Besides, the UAVs employ beamforming techniques to mitigate the interference between channels. Hence, the signal received by UAV $m$ is written as
\begin{align}
    y_{k,m}\left[ n \right] = &\nonumber {\bf w}_{k,m}^{\rm H} \left[ n \right] {\bf h}_{k,m}\left[ n \right] \sqrt{p_k\left[ n \right]} s_k\left[ n \right] + \\
    &\nonumber \sum\limits_{j=1}^M{\sum\limits_{i=1,i \ne k}^K{{\bf w}_{k,m}^{\rm H}\left[ n \right] \alpha_{i,j} {\bf h}_{i,j}\left[ n \right] \sqrt{p_i\left[ n \right] }s_i\left[ n \right] }}\\
    & + {\bf w}_{k,m}^{\rm H} \left[ n \right] {\bf n},
\end{align}
where ${\bf w}_{k,m} \left[ n \right]$ represents the unit-norm receive beamforming vector between UE $k$ and UAV $m$ during the $n$-th time slot with ${\bf w}_{k,m}^{\rm H} \left[ n \right] {\bf w}_{k,m} \left[ n \right]=1$.
Besides, ${\bf n} \sim \mathcal{CN}\left( {\bf 0}, \sigma^2{\bf I} \right)$
represents the complex vector of additive white Gaussian noise with noise variance $\sigma^2$. 
Accordingly, the resulting signal to interference plus noise ratio is calculated as 
\begin{equation}
    \Gamma_{k,m}\left[ n \right] = \frac{ \vert {\bf w}_{k,m}^{\rm H} \left[ n \right] {\bf h}_{k,m}\left[ n \right] \vert^2 p_k\left[ n \right] }{ \sum\limits_{j=1}^{M} {\sum\limits_{i=1,i \ne k}^{K} {\alpha_{i,j} \vert {\bf w}_{k,m}^{\rm H} \left[ n \right] {\bf h}_{i,j}\left[ n \right] \vert^2 p_i \left[ n \right] } } +\sigma^2 }.
\end{equation}

Thus, the offloading rate from UE $k$ to UAV $m$ during the $n$-th time slot is written as
\begin{equation}
    R_{k,m} \left[ n \right] = B \log_2 \left( 1+ \Gamma_{k,m} \left[ n \right] \right),
\end{equation}
where $B$ denotes the channel bandwidth.

\subsection{Computing Model}
In this paper, we consider different types of tasks, which can be defined as $\mathcal{Z}\triangleq\{1,2,\cdots,Z\}$.
The task of UE $k$ being accomplished in the $n$-th time slot is represented by $D_k[n] = \left(d_k[n],c_z\right)$, where $d_k[n]$ is the size of the data created by UE $k$ during time slot $n$, and $c_z$ represents the task complexity associated with the task type $z$, indicating the needed CPU processing capacity.
In practical scenarios, the task complexity $c_z$ is not always known, leading to computation uncertainty. 
This uncertainty is similar to physical world situations in which the tasks' sizes can be measured, while their processing time remains indeterminate before they are executed.
Despite the uncertainty surrounding $c_z$, we can utilize the long-term statistical information of multi-type tasks  to evaluate their task complexity, which is given by
\begin{equation}
    c_z = \hat{c}_z+\Delta\delta_z, \vert \Delta\delta_z\vert \le \varepsilon_z,
\end{equation}  
in which $\hat{c}_z$ represents the estimated task complexity of $c_z$, and $\Delta \delta_z$ is the corresponding estimation error. The permissible range of $\Delta \delta_z$ is confined within a radius of $\varepsilon_z$.
In order to schedule the task of UE $k$ during time slot $n$ with task type $z$, the matching factor between them is given by
\begin{align}
    &\zeta_{k,z}\left[ n \right] \in \{0,1\}, \forall k \in \mathcal{K}, \forall z \in \mathcal{Z},\\
    &\sum\limits_{z=1}^{Z}{\zeta_{k,z} \left[ n \right] = 1},\forall k \in \mathcal{K},
\end{align}
where $\zeta_{k,z}\left[ n \right]=1$ if the task for UE $k$ during the time slot $n$ matches the task type $z$, and $\zeta_{k,z}\left[ n \right]=0$ otherwise.

Due to the constraints in computational resources and energy, it may not be feasible to complete a task locally within the desired time frame.
In such cases, we employ a partial offloading mode in this paper and divide it into two parts.
The part with the data size of $d_k^o \left[ n \right] =  \rho_k \left[ n \right]d_k \left[ n \right]$ is executed on UAV $m$, while the remaining part with the data size of $d_k^l \left[ n \right] = \left( 1- \rho_k \left[ n \right]\right) d_k \left[ n \right]$ is processed locally,
in which $\rho_k \left[ n \right]\left( 0 \le \rho_k\left[ n \right]\le 1\right) $ is defined as the task-partition factor.

\subsubsection{Local computing}
When the task $D_k^l \left[ n \right] = \left( d_k^l \left[ n \right], c_z \right)$ is processed locally by UE $k$, the time delay can be calculated as 
\begin{equation}
    t_k^l \left[ n \right] = \frac{\sum\limits_{z=1}^Z {d_k^l \left[ n \right] c_z \zeta_{k,z}\left[ n \right]}}{f_k \left[ n \right]},
\end{equation}
where $f_k \left[ n \right]$ in [cycles/s] is UE $k$'s CPU frequency in the $n$-th time slot.

The energy consumption of local computing for UE $k$ during the $n$-th time slot is calculated as
\begin{equation}
    E_k^l \left[ n \right] = \sum \limits_{z=1}^Z {\kappa d_k^l \left[ n \right] c_z \left(f_k\left[ n \right]\right)^2  \zeta_{k,z}\left[ n \right]},
\end{equation}
in which we define $\kappa$ as the effective capacitance coefficient relying on the chip structure used.
Thus, the sum energy consumption of local computing during the $n$-th time slot is given by
\begin{equation}
    E_l \left[ n \right] = \sum \limits_{k=1}^K {E_k^l \left[ n \right]}.
\end{equation}

\subsubsection{Computation offloading}
When UE $k$ offloads $D_k^o \left[ n \right] = \left( d_k^o \left[ n \right], c_z \right)$ to UAV $m$, 
the time delay is given by
\begin{equation}
    t_k^o \left[ n \right] = \frac{d_k^o \left[ n \right]}{ \sum \limits_{m=1}^M {\alpha_{k,m}R_{k,m}\left[ n \right]}}.
\end{equation}

The energy consumption of transmission for UE $k$ during the $n$-th time slot is calculated as $E_k^o \left[ n \right] = p_k \left[ n \right] t_k^o \left[ n \right]$,
in which we define $p_k\left[ n \right]$ as UE $k$'s transmission power. Thus, the sum energy consumption of transmitting the tasks from UEs to UAVs during the $n$-th time slot
is given by 
\begin{equation}
    E_o \left[ n \right] = \sum \limits_{k=1}^K {E_k^o \left[ n \right]}.
\end{equation}

Moreover, the time delay of computing $d_k^o\left[ n \right]$ during the $n$-th time slot is given by
\begin{equation}
    t_{k}^u \left[ n \right] = \frac{ \sum \limits_{m=1}^M { \sum\limits_{z=1}^Z {\zeta_{k,z}\left[ n \right] c_z d_k^o \left[ n \right] \alpha_{k,m}} } }{ \sum \limits_{m=1}^M{\alpha_{k,m} f_{k,m}^u \left[ n \right]} },
\end{equation}
in which $f_{k,m}^u\left[ n \right]$ represents the allocated CPU frequency for UE $k$ by UAV $m$. 

Therefore, the service delay of UE $k$ is given by 
\begin{equation}
    t_k\left[ n \right] = \max \{ t_k^o\left[ n \right]+t_k^u \left[ n \right], t_k^l \left[ n \right]\} .
\end{equation}

For UAV $m$, the total energy consumption during the $n$-th time slot is denoted by
\begin{equation}
    E_{m}^u \left[ n \right] = \kappa \sum \limits_{k=1}^K {\left( \sum\limits_{z=1}^Z {\zeta_{k,z}\left[ n \right] c_z d_k^o \left[ n \right] \alpha_{k,m}}  f_{k,m}^u \left[ n \right]^2 \right)}.
\end{equation}

The UAVs' sum energy consumption of computing during the $n$-th time slot is given by 
\begin{equation}
    E_u\left[ n \right] = \sum\limits_{m=1}^M {E_m^u \left[ n \right]}.
\end{equation}

Thus, the sum weighted energy consumption in $T$ can be denoted by
\begin{equation}
    E_{\rm{total}} = \sum \limits_{n=1}^N {\left( E_l \left[ n \right] + E_o\left[ n \right]\right) + \omega \left(E_u\left[ n \right] + E_{\rm{fly}}\left[ n \right]\right)},
\end{equation}
in which $\omega$ denotes the non-negative constant weight factor.

\subsection{Problem Formulation}
Our purpose is to minimize the sum weighted energy consumption in the system by jointly configuring
the flying trajectory (i.e., ${\bf q} \triangleq \{ {\bf q}_m \left[ n \right], \forall n \in \mathcal{N},  m \in \mathcal{M} \}$),
the beamforming vector of communication symbols ${\bf w} \triangleq \{{\bf w}_{k,m}\left[ n \right], \forall n \in \mathcal{N},m \in \mathcal{M}, k \in \mathcal{K}\}$,
the task-partition factor ${\bm \rho} \triangleq \{ \rho_k \left[ n \right], \forall  k \in \mathcal{K},n \in \mathcal{N} \}$,
the matching factor between UAVs and UEs ${\bm \alpha} \triangleq \{ \alpha_{k,m}, \forall k \in \mathcal{K}, m \in \mathcal{M} \}$, the CPU frequency of UEs ${\bf f}_l \triangleq \{ f_k\left[ n \right], \forall n \in \mathcal{N},k \in \mathcal{K} \}$ and the computational resource allocation of UAVs ${\bf f}_u \triangleq \{ f_{k,m}^u \left[ n \right], \forall n \in \mathcal{N}, m \in \mathcal{M}, k \in \mathcal{K} \}$. The optimization problem is denoted by
\begin{subequations}\label{P:0}
    \begin{align}
        &\mathop {\max }\limits_{ {\bf w} ,{\bm \rho},{\bf q},{\bm \alpha},{\bf f}_l,{\bf f}_u}  E_{\rm{total}}\label{P:OB}\\
        \text{s.t.}~
        &0 \le \rho_k \left[ n \right] \le 1,\forall n \in \mathcal{N}, k \in \mathcal{K},\label{P0:eq_pho}\\
        &\sum\limits_{m=1}^M \alpha_{k,m} \le 1,\forall k \in \mathcal{K},\label{P0:eq_alpha1}\\
        &\alpha_{k,m} \in \{0,1\},\forall m \in \mathcal{M},k \in \mathcal{K},\label{P0:eq_alpha2}\\
        &\sum\limits_{z=1}^Z \zeta_{k,z}\left[ n \right]=1,\forall k \in \mathcal{K},\label{P0:eq_zeta1}\\
        &\zeta_{k,z}\left[ n \right] \in \{0,1\},\forall k \in \mathcal{K}, z \in \mathcal{Z},\label{P0:eq_zeta2}\\
        &\Vert {\bf a}_m\left[ n \right] \Vert \le a_{\rm{max}}, \forall n \in \mathcal{N},m \in \mathcal{M}, \label{P0:eq_acc}\\
        &\Vert {\bf v}_m\left[ n \right] \Vert \le v_{\rm{max}},\forall n \in \mathcal{N},m \in \mathcal{M},\label{P0:eq_v}\\
        &\Vert {\bf q}_i\left[ n \right] - {\bf q}_j\left[ n\right] \Vert^2 \ge d_{\rm {dim}}^2,\forall i,j \in \mathcal{M}, i \ne j,\label{P0:eq_diatance}\\
        &0\le p_k\left[ n \right] \le p_{k,\rm{max}},\forall n \in \mathcal{N}, k \in \mathcal{K},\label{P0:eq_pk}\\
        &0\le f_k\left[ n \right] \le f_{k,\rm{max}},\forall n \in \mathcal{N}, k \in \mathcal{K},\label{P0:eq_fk}\\
        &0\le f_{k,m}^u\left[ n \right] \le f_{u,\rm{max}},\forall k \in \mathcal{K}, n \in \mathcal{N}, m \in \mathcal{M}, \label{P0:eq_fu}\\
        &0\le \sum \limits_{k=1}^K {\alpha_{k,m}\left[ n \right] f_{k,m}^u\left[ n \right]} \le f_{u,\rm{max}}, \forall m \in \mathcal{M},n \in \mathcal{N},\label{P0:eq_fu_sum}\\ 
        &\mathop {\max }\limits_{\vert \Delta \delta_z \vert, \Vert \Delta h_{k,m}\left[ n \right] \Vert} t_k\left[ n \right] \le \delta, \forall k \in \mathcal{K},n \in \mathcal{N},\label{P0:eq_delay}\\
        &\Vert \Delta {\bf h}_{k,m}\left[ n \right] \Vert \le \varepsilon_{k,m},\forall m \in \mathcal{M}, n \in \mathcal{N}, k \in \mathcal{K},\label{P0:eq_hrob}\\
        &\vert \Delta \delta_z \vert \le \varepsilon_z,\forall z \in \mathcal{Z},\label{P0:eq_crob}
    \end{align}
\end{subequations}
where $p_{k,\rm{max}}$ is the maximum transmission power, $f_{k,\rm{max}}$ and $f_{u,\rm{max}}$ are the maximum CPU frequency of UEs and UAVs, respectively. $v_{\rm{max}}$ is the maximum speed when UAVs fly and $a_{\rm{max}}$ is the maximum UAV acceleration.
Constraint \eqref{P0:eq_pho} represents the task offloading ratio.
Constraint \eqref{P0:eq_alpha1} and constraint \eqref{P0:eq_alpha2} reflect that the UE is limited to connecting with a single UAV at most.
Constraint \eqref{P0:eq_zeta1} and constraint \eqref{P0:eq_zeta2} reflect that the task only belongs to one task type.
Constraint \eqref{P0:eq_acc} and constraint \eqref{P0:eq_v} are UAVs' speed and acceleration limitations.
Constraint \eqref{P0:eq_diatance} is the minimum safe diatance limitation between UAVs.
Constraint \eqref{P0:eq_pk} is the transmission power requirements of UEs.
Constraints \eqref{P0:eq_fk}-\eqref{P0:eq_fu_sum} are the computation resource constraints of UEs and UAVs.
Constraint \eqref{P0:eq_delay} denotes the computing delay requirements.
Constraint \eqref{P0:eq_hrob} and constraint \eqref{P0:eq_crob} are the robust constraints related to communication and computation.

\section{MAPPO-Based Algorithm for Robust Offloading and Trajectory Optimization}
It can be derived that problem \eqref{P:0} belongs to a complicated nonconvex problem since it includes a highly nonconvex objective function and discrete variables.
Moreover, the uncertainties and dynamic features of the environment, caused by the time-varying channel conditions and diverse task types, invoke a significant challenge for traditional offline optimization techniques.
To achieve a real-time online decision-making for configuring heterogeneous resources, DRL has been proposed to determine the optimal joint configuration.
However, the training scenarios featuring high-dimensional action and state spaces is intractable to handle for single-agent DRL algorithms.
In addition, the latency cost will rise as a result of the frequent synchronization of state information between network entities.
Thus, we propose a training framework based on MAPPO for the multi-UAV-assisted MEC network, which enables the collaboration and distribution of multiple policy types to jointly determine the optimization variables.
\subsection{Modeling of Multi-agent MDP}
In this network, there are multiple UAVs and UEs, and the optimization problem exhibits distributional characteristics of real-world scenarios.
Hence, the problem can be expressed as a multi-agent Markov Decision Process (MDP).
Typically, MDP involves three essential components: a reward function $\mathcal{R}$, a state space $\mathcal{S}$ and a action space $\mathcal{A}$. 
In a multi-agent system, each agent $i \in \mathcal{I} \triangleq \{1,2,\dots,I \}$ makes observations denoted by $o_n^i$ at time step $n$.
And all agents' partial observations are combined to obtain the global state $s_n$.
To facilitate decision-making and achieve near-optimal solutions, we propose to decompose the general policy into two policies, one for UE agents and another for UAV agents.
Thus, we have $I = K + M$.
Besides, the global state space $\mathcal{S}= \mathcal{O}_1 \times \dots \times \mathcal{O}_I$ is the Cartesian product of all observation spaces $\mathcal{O}_i$ while the action space $\mathcal{A} = \mathcal{A}_1 \times \dots \times \mathcal{A}_I$ is the Cartesian product of all action spaces $\mathcal{A}_i$ for all agents.
These two types of policies can be presented as follows:
\subsubsection{UE agent}
UE agent emphasizes the local computing for UEs and configures task offloading accordingly.
The index set of UE agents can be given by $I_1 \triangleq \{1,2,\dots,K\}$.
Besides, observing the locations of both themselves and UAVs, as well as the task-related information, is necessary to determine the association to UAVs and the offloading proportion.

{\bf{Observation:}} The observation of UE agent is denoted as:
\begin{equation}
    o_n^k = \{k, {{\bf{q}}_m\left[ n \right]}, D_k\left[ n \right], {{\bf{u}}_k}, \zeta_{k,z}\left[ n \right], \forall m \in \mathcal{M}, \forall z \in \mathcal{Z} \},
\end{equation}
where each UE is only capable of accessing its own location information through a positioning service while the information of all UAVs can be accessed by UEs since UAVs act as servers.
To minimize the energy consumption during computing, the CPU frequency $\hat{f}_k\left[ n \right]$ can be simply estimated by using the dynamic voltage frequency scaling technology, which can be expressed by the following equation $\hat{f}_k\left[ n \right] = \min\{ f_{k,\rm{max}}, \frac{\sum_{z=1}^{Z}{\rho_k\left[ n \right] d_k\left[ n \right] \zeta_{k,z}\left[ n \right]}}{t_k\left[ n \right]} \}$.

{\bf{Action:}} The action of UE agent should reflect the decision variables, and therefore can be given by
\begin{equation}
    a_n^k = \{\alpha_{k,m}, \rho_k\left[ n \right], \forall m \in \mathcal{M} \}.
\end{equation}

For the constraints \eqref{P0:eq_alpha1} and \eqref{P0:eq_alpha2}, $\hat{m}_k = arg \mathop {\max }\limits_{m} \{ \hat{\alpha}_{k,m}, \forall m \in \mathcal{M} \}$ is selected as the associated UAV of UE $k$, and $\hat{\alpha}_{k,m}$ denotes the output of the policy model.
Besides, $\hat {\rho}_k \left[ n \right] \le 0$ represents the case of fully local computing. Hence, we can map the range of output $\hat{\rho}_k \left[ n \right]$ into $\left[ -\varepsilon, 1\right]$ where $\varepsilon >0 $.

{\bf{Reward:}} To design an effective UE agent policy, its reward function should include both the objective and the penalty for not meeting the latency requirements.
Furthermore, the decomposition of the energy consumption of UEs and their associated UAVs for each individual UE also needs to be taken into account.
As a result, the reward can be denoted as
\begin{equation}
    r_n^k = -E_k^{\omega} \left[ n \right] P_{T,k}^u \left( n \right),
\end{equation}
in which
\begin{equation}
  \nonumber  E_k^\omega \left[ n \right] = E_k^l\left[ n \right]+E_k^o\left[ n \right] + \omega \sum_{m=1}^{M}{\alpha_{k,m}\left( E_m^u \left[ n \right] + E_m^{\rm{fly}}\left[ n \right] \right)}
\end{equation}
represents UE $k$'s weighted energy consumption.
$P_{T,k}^u \left( n \right)$ is calculated as 
\begin{equation}
    P_{T,k}^u \left( n \right) = P\left( t_k\left[ n \right] , t_k^{\rm{max}}\left[ n \right],t_k^{\rm{max}}\left[ n \right] \right),
\end{equation}
where
\begin{equation}
    P\left(r,p,q\right) = 2 - \exp{\left(-\left\lceil \left(r-p\right)/q \right\rceil^+\right)}.
\end{equation}
\subsubsection{UAV agent} The CPU frequency allocation for UEs served by UAVs, as well as the control of UAVs' flying speed, should be managed by UAVs.
The index set of UAV agents can be given by $I_2 \triangleq \{ K+1,K+2,\dots,K+M \}$.
The observation, action and reward of the UAV agent can be illustrated as:

{\bf{Observation:}} Every UAV is capable of acquiring both the computation offloading information and the location of UEs served by itself.
Hence, the observation of each agent can be given by
\begin{align}
    \nonumber o_n^{K+m} = &\{m,{\bf{u}}_k\left[ n \right], {\bf{q}}_m\left[ n \right], {\bf{q}}_{-m}\left[ n \right],  \\
    & \rho_k\left[ n \right],D_k\left[ n \right],\forall k \in\mathcal{K}_m \},
\end{align}
in which we define $\mathcal{K}_m$ as UEs served by UAV $m$, and $-m$ as the indexes in set $\mathcal{M} \backslash m$.

{\bf{Action:}} The UAVs can improve the fairness of UEs by deciding their movement and allocating the CPU frequency to process UEs' tasks.
Hence, the actions of UAV agents are denoted as
\begin{equation}
    a_n^{K+m} = \{ {\bm{a}}_m\left[ n \right], {\bf{w}}_{k,m}\left[ n \right], f_{k,m}\left[ n \right],\forall k \in \mathcal{K}_m \}.
\end{equation}

We define $\hat{\bm{a}}_m\left[ n \right] = \left[ \Vert {\bm{a}}_m\left[ n \right] \Vert, \phi_m\left[ n \right] \right]$ as the output acceleration where $\phi_m\left[ n \right]$ denotes the  angular acceleration.
Besides,  a vector with a length of $K + 1$ can be used to represent the available computation resources of a UAV and the proportion of resources allocated to each UE.
If UAV $m$ doesn't serve UEs, it will be multiplied by zero.
Thus, the estimated value of CPU frequency can be considered as a representation of the action taken.

{\bf{Reward:}} The UAV $m$ should balance the energy consumption and the distance to UEs to improve the channel gain and fairness simultaneously.
Besides, it's important to take into consideration the penalties induced by collisions and objects flying out.
Thus, the reward can be denoted as follows

\begin{align}
    \nonumber r_n^m =& - \left(\right. \kappa_1 \tilde{E}_m\left[ n \right] + \kappa_2 P\left(\right. \Vert {\bm q}_m\left[ n \right] - \\
    &\frac{1}{\vert \mathcal{K}_m \vert}\sum_{k \in \mathcal{K}_m}{\alpha_{k,m}{\bm u}_k \left[ n \right]}\Vert, d_{\rm{th}},X \left.\right)\left. \right)P_{n,T}^m P_{n,o}^m P_{n,c}^m,
\end{align}
in which $\kappa_1$ and $\kappa_2$ are both defined as the adjustment factors, $X$ represents the width of square service region, and $d_{\rm{th}}$ represents the threshold distance between UAVs and UEs.
We define $\tilde{E}_m\left[ n \right]$ as the weighted average energy consumption, which is modeled as
\begin{align}
    \nonumber \tilde{E}_m\left[ n \right] = &\frac{1}{\vert \mathcal{K}_m \vert} \sum_{k \in \mathcal{K}_m}\left[\right.\alpha_{k,m}\left( E_k^o\left[ n \right] + E_k^l \left[ n \right] \right)\\
    & +\varpi \left( E_m^u\left[ n \right] + E_m^{\rm{fly}}\left[ n \right] \right)\left.\right],
\end{align}
where $\varpi$ is the adjusting factor.
The penalties are respectively given by $P_{n,T}^m$, $P_{n,o}^m$ and $P_{n,c}^m$.
Specifically, the penalty for not meeting the latency requirements of UEs served by UAV $m$ is represented by
\begin{equation}
    P_{n,T}^m = \frac{1}{\vert \mathcal{K}_m \vert} \sum_{k \in \mathcal{K}_m}{P\left( \alpha_{k,m}t_k\left[ n \right], t_k^{\rm{max}\left[ n \right]}, t_k^{\rm{max}\left[ n \right]}\right)},
\end{equation}
the penalty for flying out of the service region is given by
\begin{equation}
    P_{n,o}^m = 1+ \frac{1}{v_{\rm{max}}}\Vert {\bm q}_m \left[ n \right] - {\rm{clip}}\left( {\bm q}_m \left[ n \right], 0 , X \right)\Vert,
\end{equation}
and the penalty for not maintaining a safe distance between UAVs is represented by
\begin{equation}
    P_{n,c}^m = \sum_{j=1,j \ne m}^{M}{P\left( d_{\rm{min}}, \Vert {\bm q}_m -{\bm q}_j \Vert, d_{\rm{min}} \right)}.
\end{equation}

\begin{figure}[t]
    \captionsetup{justification=raggedright,singlelinecheck=false}
    \includegraphics[width=\columnwidth]{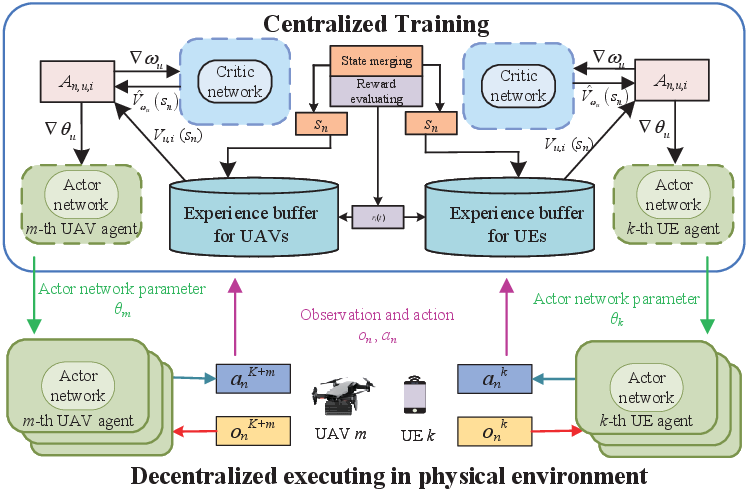}
    \caption{The training framework of b-MAPPO.}
    \label{fig:MAPPO}
\end{figure}
\subsection{MAPPO-based DRL Training Framework}
On-policy DRL approaches are widely known for their stable training performance and efficient use of computational resources, allowing devices to allocate more resources to other significant functions.
Thus, MAPPO is designed to train the multi-agent policies that can achieve high performance on the target task while maintaining training stability.
MAPPO is an on-policy MADRL algorithm based on the actor-critic framework, which has shown excellent results on diverse tasks.
In MAPPO, the actor network $\theta_u$ expresses actions, the critic network $\omega_u$ evaluates the state-value function, and the shared policy of UE or UAV agents is represented by $\pi_{\theta_u}$.

For easy deployment in distributed networks,  the centralized training and decentralized executing framework is considered as shown in Fig. \ref{fig:MAPPO}.
Under this framework, UEs and UAVs perform computation offloading based on the actions provided by their respective actor networks and send their experiences to the training center.
Then, the global environment state is evaluated by the observations of agents, the buffers are updated, and the prediction values are obtained.
After updating the actor and critic networks, the parameters of the actor network are downloaded to UAVs and UEs. 
Moreover, note that the network parameters are shared among the homogeneous agents.

In this framework, the state-value function of the $u$-th type of agents is represented by
\begin{equation}
    V_{u,i}^{\pi}\left(s_n,\theta_u\right) = \mathbb{E} \{ \sum_{l=0}^{\infty}{\gamma_u^l \mathcal{R}_{u,i}\left( s_{n+l},a_{n+l}|s_n=s,\pi \right)} \},
\end{equation}
in which $\mathbb{E}\{\cdot\}$ represents the expectation operation, $\mathcal{R}_{u,i}$ represents the reward function of the $i$-th agent of the $u$-th type of agent, $a_n$ is the action of all agents, $\pi$ is the policy of agents, and the discount factor $\gamma_u$ represents the significance of forthcoming rewards for all agents.
The action-value function can be denoted as
\begin{equation}
    Q_{u,i}^\pi\left(s_n,a_n\right) = \mathbb{E}\{ \sum_{l=0}^{\infty}{\gamma_u^l \mathbb{R}_{u,i}\left( s_{n+l},a_{n+l}\right)|s_n = s,a_n=a,\pi } \}.
\end{equation} 

On this basis, to calculate the advantage value of each action which can be used to update the strategy,
the advantage function can be denoted as $A_{n,u,i} = Q_{u,i}^\pi\left(s_n,a_n\right) - V_{u,i}^\pi\left(s_n\right)$, and it can be evaluated as $\hat{A}_u\left(s_n\right) = \sum_{l=0}^{\infty}{\left(\gamma_u \lambda\right)^l \left( r_{n+l}+\gamma_u V_u\left( s_{n+l+1} \right) - V_u\left(s_n\right) \right)}$ by utilizing the state-value $V_u\left(s_n\right)$.
It should be noted that we make use of the Generalized Advantage Estimation (GAE) to estimate the advantage function, and $\lambda$ represents the GAE factor, which plays a significant role in balancing the bias and variance of the rewards.
Besides, $\delta_n = \left( r_n+\gamma_u V_u\left( s_{n+1} \right) - V_u\left(s_{n+l}\right) \right)$ denotes the temporal-difference error.
Denoting $\hat{V}_{\omega_u}\left(s_n\right)$ as the state-value function estimated by the critic network, we can use the following loss function to update the critic network:
\begin{equation}
    J\left( \omega_u \right) = \frac{1}{2}\left[ \hat{V}_{\omega_u}\left(s_n\right) - V_u\left(s_n\right) \right]^2.\label{J_omega}
\end{equation}

For the actor networks, the clipping factor $\varepsilon$ is introduced into MAPPO algorithm in order to limit the update ratio of policy.
Thus, the actor network's loss function is calculated as
\begin{align}
    \nonumber J \left( \theta_u \right) = &\mathbb{E}\{ \min \left[ {\rm{clip}}\left( \frac{\pi_{\theta_u}\left(a_n|s_n\right)}{\pi_{\theta'_u}\left(a_n|s_n\right) }, 1-\varepsilon, 1+\varepsilon \right) \hat{A}_u\left(s_n\right), \right.\\
    &\left. \frac{\pi_{\theta_u}\left(a_n|s_n\right)}{\pi_{\theta'_u}\left(a_n|s_n\right) }\hat{A}_u\left(s_n\right) \right] + \psi S_{n,u}\},\label{J_theta}
\end{align} 
in which $\theta'_u$ denotes the parameters of the old policy. The update ratio is denoted by $\frac{\pi_{\theta_u}\left(a_n|s_n\right)}{\pi_{\theta'_u}\left(a_n|s_n\right) }$, and the policy entropy of the degree of exploration is represented as $\psi S_{n,u}$. 
Thus, we can utilize the gradients $\nabla\theta_u = \frac{\partial J\left(\theta_u\right)}{\partial \theta_u}$ and $\nabla \omega_u = \frac{\partial J\left( \omega_u \right)}{\partial \omega_u}$ to update the actor and critic networks.
\subsection{Beta Policy}
In policy-based DRL algorithms, the Gaussian distribution has been widely utilized to model the output of actor networks.
However, this distribution is unbounded, whereas many actions have predefined lower and upper limits.
As a result, these actions must be constrained within these boundaries, which in turn creates the boundary effects that negatively impact performance \cite{beta}.
In addition, setting a small initial variance in the Gaussian distribution to reduce boundary effects can limit the exploration ability of the network by concentrating the probability density too much.
Conversely, setting a larger variance can lead to the values of actions being clipped at the boundaries, thereby reducing exploration.
Therefore, we introduce the Beta distribution into the actor network's output.
The Beta distribution with respect to $x$ is denoted as \cite{set1}
\begin{equation}
    f\left(s,\tau,\zeta\right)=\frac{\Gamma\left( \tau+\zeta \right)}{\Gamma\left(\tau\right)\Gamma\left( \zeta \right)}s^{\tau-1}\left( 1-s \right)^{\zeta-1}\label{eq_beta}.
\end{equation}

It can be derived that \eqref{eq_beta} has a bounded domain, and thus it is adaptable to the actions that have double boundaries.
Moreover, it also facilitates the algorithm to conduct more uniform exploration during the early stage of training.
Correspondingly, compared to the Gaussian distribution, the Beta distribution typically exhibits higher probability density near its boundaries.
Based on the Beta distribution, we summarize the b-MAPPO training framework and the pseudocode is shown in Algorithm 1.
\begin{algorithm}[t]
    \caption{Proposed b-MAPPO training framework}
    \label{b-MAPPO}
    \begin{algorithmic}[1]
        \STATE{Initialize the maximum training episodes Mt, the episode length epi and the PPO epochs epc.}
        \STATE{
            Initialize critic networks $\omega_i$, actor networks $\theta_i$ of UEs and UAVs, $\forall i \in \{1, 2\}$;}
        \FOR{ep=1 to Mt}
            \FOR{n=1 to epi}
                \STATE{Obtain observations $o_n^i$ from the environment, $\forall i \in \mathcal{I}_1$;}
                \STATE{Execute actions $a_n^i$, $\forall i \in \mathcal{I}_1$;}
                \STATE{Obtain observations $o_n^i$ from the environment, $\forall i \in \mathcal{I}_2$;}
                \STATE{Execute actions $a_n^i$, $\forall i \in \mathcal{I}_2$;}
                \STATE{The UEs and UAVs send the observations and actions to the execution center and the center measures the rewards $r_n^i$;}
            \ENDFOR
            \STATE{Calculate log-probability $p_n^i, \forall i \in \mathcal{I}, n \in \{ 1,\cdots, {\rm{epi}} \}$;}
            \STATE{Summarize the transitions ${\rm{tre}}_n^i = \{ o_n^i,a_n^i,r_n^i,s\left( n \right),p_n^i,\forall i \in \mathcal{I},n \in \{ 1,\cdots,{\rm{epi}} \} \}$ in buffers;}
            \FOR{epo = 1 to epc}
                \FOR{ agents $i \in \mathcal{I}$}
                    \STATE{Adjust $\omega_i$ and $\theta_i$ according to \eqref{J_omega} and \eqref{J_theta};}
                \ENDFOR
            \ENDFOR
        \ENDFOR
    \end{algorithmic}
\end{algorithm}
\subsection{Complexity Analysis}
In this subsection, we analyze the computational complexity of the proposed b-MAPPO algorithm.
In this framework, for Multi-Layer Perceptron (MLP), the computational complexity of the $i$-th layer can be expressed as $\mathcal{O}\left( L_{i-1}L_i+L_i L_{i+1} \right)$, in which the number of neurons in $i$-th layer is defined as $L_i$.
Thus,  the computational complexity of an $I$-layer MLP can be denoted as $ \mathcal{O}\left( \sum_{i=2}^{I-1}{L_{i-1}L_i+L_i L_{i+1}} \right) $.
In our algorithm, the actor networks have one MLP each, and the critic networks have one MLP for value output and two encoders for two types of agents.
Besides, due to the fact that in a decision step,  the agents are capable of computing their actor networks in parallel, the complexity can be represented as $ \mathcal{O}\left( \sum_{i=2}^{I-1}{L_{i-1}L_i+L_i L_{i+1}} \right) $.
Thus, with all Mt episodes, the time complexity of the training algorithm is $\mathcal{O}\left( {\rm{Mt}} \left( {\rm{epi}}\left( \sum_{i=2}^{I-1}{L_{i-1}L_i+L_i L_{i+1}} \right) \right) \right)$.

\section{ NUMERICAL RESULTS}
This section presents simulation experiments to illustrate the effectiveness of the proposed b-MAPPO training framework in a multi-UAV-assisted MEC network.
In the simulation, we set UAVs' service region to be a square-shaped area with side length of 1000 m, where UEs  are randomly and uniformly distributed and the initial horizontal locations of UAVs are randomly set with $x,y \in \left[0,1000\right]$ m.
The number of UEs is $K=20$ and the number of UAVs is $M=5$.
The size of task is uniformly distributed in [$D_{\rm{min}}$, $D_{\rm{max}}$], in which $D_{\rm{min}}$ and $D_{\rm{max}}$ are set to be 3.5Mb and 4.5 Mb as default \cite{Video}.
The mean number of cycles per bit for the tasks is $c_z \in [500,1500]$.
The confidence interval is set as 95\%.
To algorithm setup, we use the value normalization and all the rewards are forced into [-5, 5].
The maximum training episodes are Mt $=$ 300 episodes, the episode length epi, which represents the $T$, is 200 steps, the discount factor is $\gamma_u=0.98$, the learning rate is 0.0005,  and the optimizer we used is Adam. Other parameter settings of the simulation are summarized in Table \ref{table_1}, according to prior work \cite{Robust7}, \cite{set1}, \cite{set2}.
\renewcommand{\arraystretch}{1.5}
\begin{center}
    \begin{table}[h]
        \captionsetup{labelfont={color=black}}
        \caption{SIMULATION PARAMETERS}
        \begin{tabularx}{0.47\textwidth}{XX|XX}
            \hline
             Parameters & Values & Parameters & Values\\
            \hline
            $Z$ & 5 & $H$ & 200 m\\
            $\varepsilon_{k,m}$ & 0.05 & $\varepsilon_z$ & 20 \\
            $B$ & 10 MHz & $\delta$ & 1.0 s\\
            $p_{k,\rm{max}}$ & 0.5 W & $f_{k,\rm{max}}$ & 1 GHz\\
            $f_{d,\rm{max}}$ & 10 GHz & $A$ & 4\\
            $a_{\rm{max}}$ & 5 $\rm{m/s^2}$ & $v_{\rm{max}}$ & 20 m/s\\ 
            $P_1$ & 59.03 W & $P_2$ & 79.07 W\\
            $U_{\rm{tip}}$ & 120 m/s & $A_0$ & 0.5030 $\rm{m/s^2}$\\
            $v_0$ & 3.6 m/s & $s$ & 0.05\\
            $\sigma^2$ & -85 dBm & $\varsigma$ & 10\\
            \hline
        \end{tabularx}
        \label{table_1}
    \end{table}
\end{center}

We compare the performance of the proposed b-MAPPO algorithm with the following benchmarks:
\begin{itemize}
	\item {\bf{Pure-MAPPO:}} The method is the original MAPPO algorithm without the use of the Beta distribution-based improvement mechanism, and it shares the same reward function, action space, and state space as the proposed algorithm \cite{22}.
	\item {\bf{MADDPG (Multi-Agent Deep Deterministic Policy Gradient):}} This method is currently popular and reliable multi-agent reinforcement learning algorithm adopted by works such as \cite{UAV6} and \cite{24}. It consists of dual actor networks and dual critic networks, where the output of the actor network serves as the action values, which are then added with certain exploration noise, and the action-value function is evaluated by the critic network.
	\item {\bf{Greedy:}}  This algorithm greedily selects the UAV trajectory, the task partition, and the computation and communication resource allocation in the $n$-th time slot to minimize the energy consumption, based on the current knowledge.
	\item {\bf{DRL+CVX:}} This algorithm uses the CVX solver for obtaining the optimal task partition variable, and uses our b-MAPPO to find the near-optimal UAV trajectory and the allocation of computation and communication resources, similar to \cite{CVX}.
\end{itemize}

\begin{figure}
    \captionsetup{justification=raggedright,singlelinecheck=false}
    \includegraphics[width=\columnwidth]{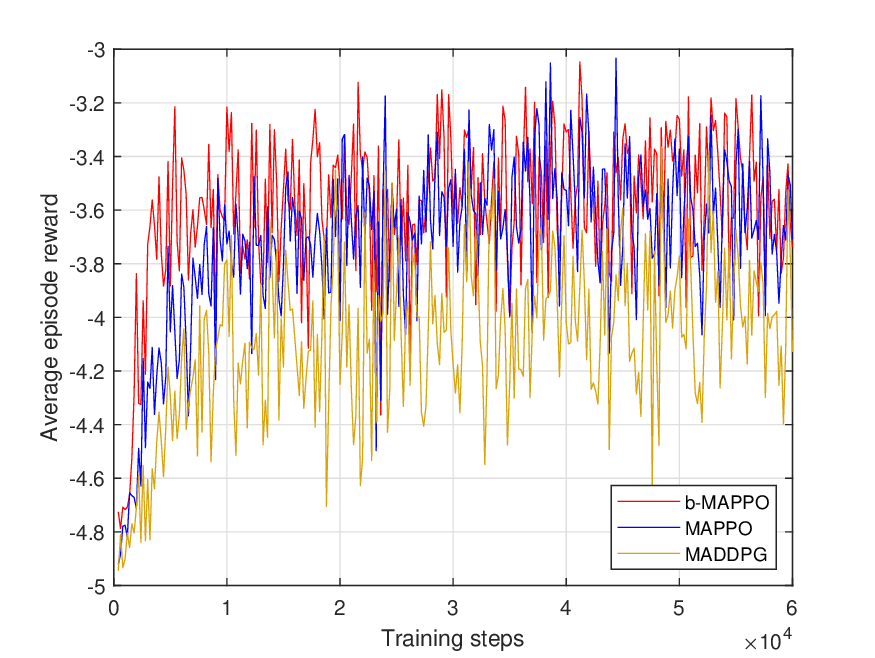}
    \caption{Convergence versus UE agents.}
    \label{fig:user_averagereward}
\end{figure}

\begin{figure}
    \captionsetup{justification=raggedright,singlelinecheck=false}
    \includegraphics[width=\columnwidth]{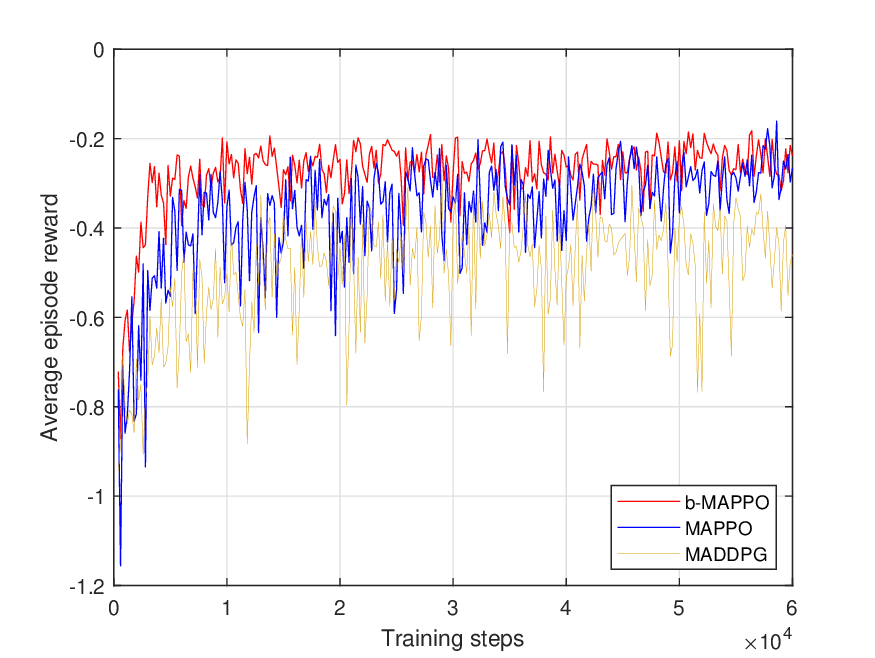}
    \caption{Convergence versus UAV agents.}
    \label{fig:uav_averagereward}
\end{figure}
In Fig. \ref{fig:user_averagereward} and Fig. \ref{fig:uav_averagereward}, we demonstrate the convergence performance of the proposed b-MAPPO algorithm compared to other benchmark methods.
With the number of training iterations growing larger, the reward obtained by all the algorithms gradually improves, indicating the efficacy of the MADRL algorithms for computation offloading.
Moreover, it is obvious that the b-MAPPO algorithm achieves the highest reward and exhibits a faster convergence rate compared to the Pure-MAPPO with Gaussian distribution and MADDPG algorithms.
Thus, it proves that the Beta distribution has a better effect than Gaussian distribution in our network.
Besides, we can find from Fig. \ref{fig:user_averagereward} that the reward received by UE agents shows a gradual improvement over time and the proposed b-MAPPO scheme achieves an average episode reward of approximately -3.05, which is the highest value observed in the experiment.
Fig. \ref{fig:uav_averagereward} illustrates how UAV agents adjust their policy to achieve a satisfactory trade-off between the positioning of the served UEs and the energy consumption.

\begin{figure}
    \centering
    \includegraphics[width=\columnwidth]{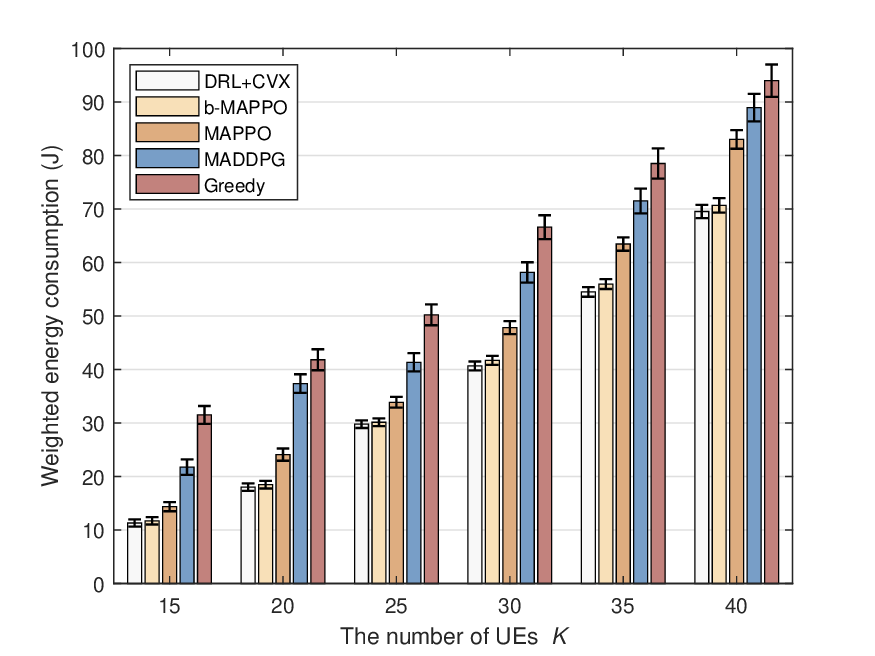}
        \caption{Performance comparison versus different numbers of UEs.}
    \label{fig:user_energy}
\end{figure}
Fig. \ref{fig:user_energy} provides a comparison of the weighted energy consumption for different numbers of UEs. 
The results indicate that the DRL-based algorithms perform better than the Greedy algorithm since the DRL-based algorithms  can adapt to uncertain environments by continuously interacting with the environment, while the Greedy algorithm is more prone to getting stuck in local optimal solution. 
Furthermore, the b-MAPPO algorithm outperforms MAPPO and MADDPG algorithms, and there is still a significant performance gap between the MADDPG-based and MAPPO-based algorithms. 
Besides, our algorithm shows minimal difference compared to the DRL+CVX algorithm with a lower computational complexity. 
Additionally, as the number of UEs increases, the weighted energy consumption also increases. 
This is because more UEs need more computation and communication resources and the increase in signal interference between UEs results in slower transmission rates.

\begin{figure}
    \centering
    \includegraphics[width=\columnwidth]{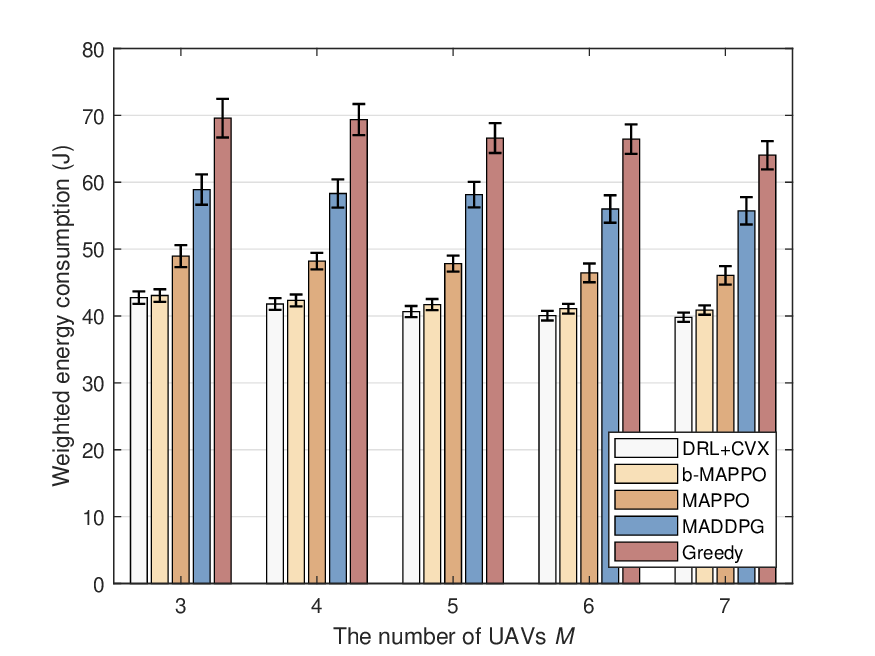}
    \caption{The performance comparison versus different numbers of UAVs.}
    \label{fig:UAV_energy}
\end{figure}
Fig. \ref{fig:UAV_energy} compares the performance of five schemes versus different numbers of UAVs under $K=30$ UEs. 
As the number of UAVs increases, there is a noticeable trend of the reduced weighted energy consumption.
This phenomenon can be explained by the fact that a larger pool of computational resources becomes available with the growth in the number of UAVs.
This allows the agents to achieve a better trade-off between the computing load on UAVs and UEs, thereby reducing the overall weighted energy consumption.
Moreover, our b-MAPPO scheme outperforms MAPPO, MADDPG, and Greedy in all scenarios, and it shows only a small performance gap compared to the DRL+CVX algorithm.

\begin{figure}
    \centering
    \includegraphics[width=\columnwidth]{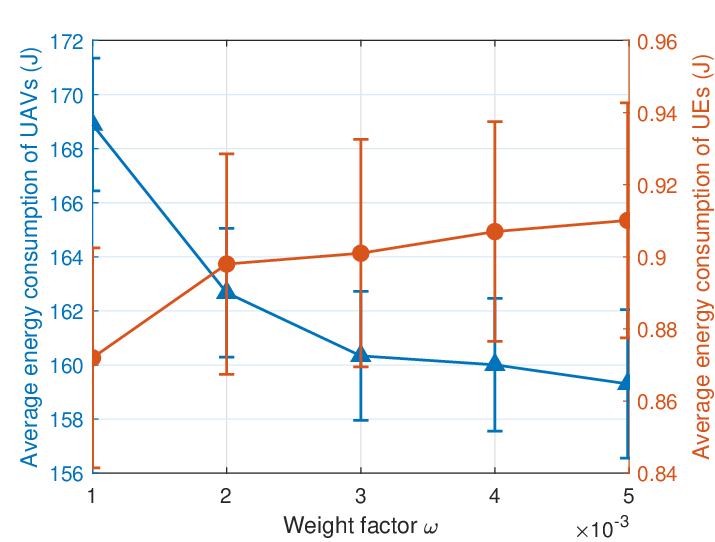}
    \caption{The influence of weight factor $\omega$ on energy consumption.}
    \label{fig:weight}
\end{figure}
Fig. \ref{fig:weight} displays the average energy consumption of UAVs and UEs for different weight factors $\omega$ to investigate the relationship on energy consumption between UEs and UAVs.
As observed, with the growth of the weight factor $\omega$, the energy consumption of the UE slowly increases, while the UAV's energy consumption decreases.
This is attributed to the trade-off function of $\omega$ on the objective, which changes the relative importance of energy consumption for both UEs and UAVs, leading to corresponding changes in policies.
The relative importance of energy consumption can be evaluated based on factors such as power capacity.

\begin{figure}
    \centering
    \includegraphics[width=\columnwidth]{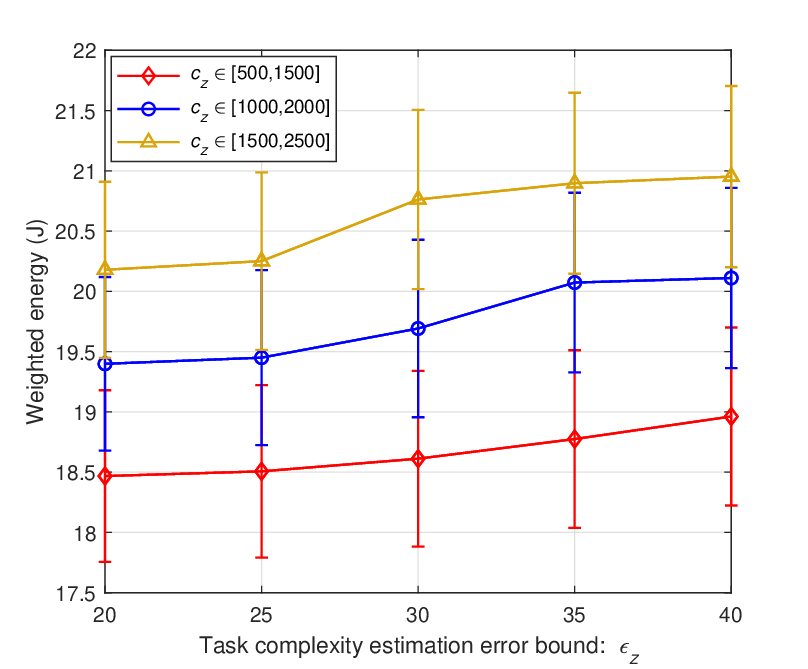}
    \caption{The performance versus different task complexity estimation error bounds under different task complexity.}
    \label{fig:computation}
\end{figure}
Fig. \ref{fig:computation} illustrates the impact of task complexity estimation error bounds on performance, with different distributions of task complexity $c_z$ across intervals.
The results indicate that wider intervals of task complexity $c_z$ lead to higher weighted energy consumption.
This can be attributed to the fact that a larger $c_z$ requires more computational workload for the task, resulting in greater energy consumption, even under the same estimation error bound.
Moreover, as the estimation error bound increases, the energy consumption also increases. The reason is that a larger error bound leads to greater uncertainty in computation.

\begin{figure}
    \centering
    \includegraphics[width=\columnwidth]{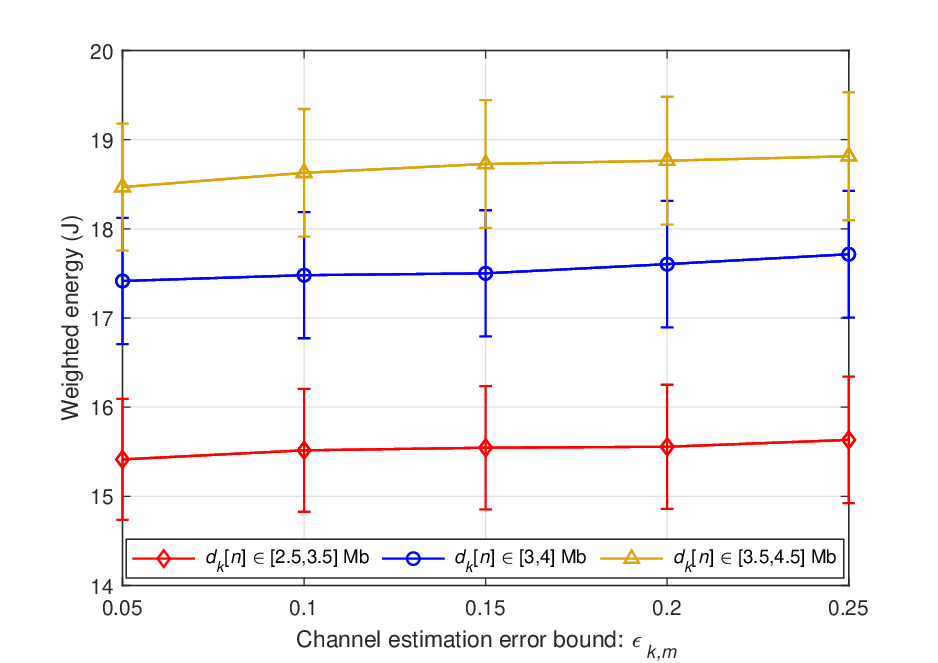}
    \caption{ The performance versus different channel estimation error bounds under different task sizes.}
    \label{fig:channel}
\end{figure}
In Fig. \ref{fig:channel}, the impact of various channel estimation error bounds on different data sizes is illustrated.
It is evident that as the channel estimation error bound grows, the system's energy consumption also increases.
The reason is that a larger error bound $\varepsilon_{k,m}$ implies higher communication uncertainty, which leads to a more significant performance degradation for a given data size.
Additionally, the weighted energy consumption increases with the growth of the data size.
The reason is that the larger data size requires more resources for transmission and computation, leading to the growth of the system's weighted energy consumption.

\begin{figure}
    \centering
    \includegraphics[width=\columnwidth]{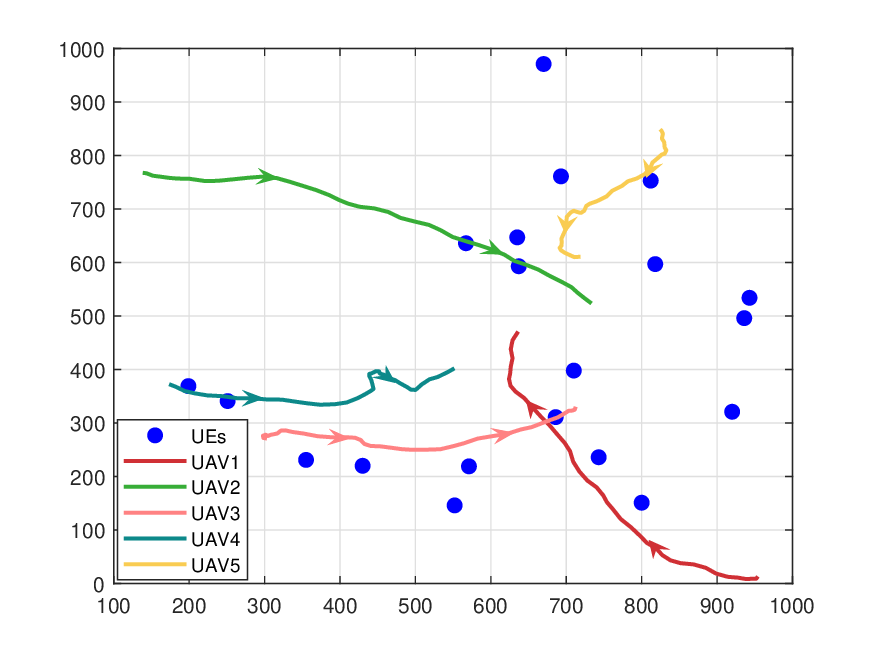}
    \caption{ The example of trajectories of UAVs under $K=20$ and $M=5$.}
    \label{fig:tr}
\end{figure}
In Fig. \ref{fig:tr}, we demonstrate the trajectories of UAVs.
It is evident that UAVs have the capability to identify regions with a higher concentration of UEs and adjust their positions accordingly based on UE distribution. 
Additionally, the figure portrays how the reward mechanism can assist UAVs in discovering a relatively equitable area for UEs and then move gradually to conserve flying energy consumption. 

\section{CONCLUSION}
In this paper, considering both the communication and computation uncertainties, we proposed a robust computation offloading scheme for the multi-UAV-assisted MEC networks.
We formulated a system energy consumption minimization problem by the joint optimization of the beamforming vector, the task-partition factor, the flying trajectory, the matching factor, the CPU frequency of UEs and UAVs.
In order to address the optimization problem, a b-MAPPO distribution framework was developed to achieve an optimal learning strategy efficiently.
Extensive numerical results showed that the proposed scheme outperforms the benchmarks in reducing energy consumption.
In our future work, we will further investigate the scenario in which different types of tasks are allowed to use different offloading rates.
\bibliographystyle{IEEEtran}
\bibliography{IEEEabrv,refs}

\end{document}